\documentclass[prl,amsmath,twocolumn, showpacs, superscriptaddress,10pt]{revtex4-1}

\usepackage{amsmath}
\usepackage{hyperref}
\usepackage{graphicx}

\expandafter\let\csname equation*\endcsname=\relax
\expandafter\let\csname endequation*\endcsname=\relax

\usepackage{enumerate}
\usepackage{mathbbol}
\usepackage{amsfonts}
\usepackage{color}
\definecolor{Blue}{rgb}{0.00, 0.00, 1.00}
\definecolor{Red}{rgb}{1.00, 0.00, 0.00}

\def\XXint#1#2#3{{\setbox0=\hbox{$#1{#2#3}{\int}$}
     \vcenter{\hbox{$#2#3$}}\kern-.5\wd0}}

\begin{document}

%\title{Random walks and L\'evy flights with stochastic resetting} 
\title{The Emergence of Weak Criticality in SOC systems}

\author{Lorenzo \surname{Palmieri}}
\affiliation{Centre for Complexity Science and Department of Mathematics, Imperial College London, South Kensington Campus, SW7 2AZ, UK;}
\author{Henrik Jeldtoft \surname{Jensen$^{1,}$}}
\affiliation{Institute of Innovative Research, Tokyo Institute of Technology, 4259, Nagatsuta-cho, Yokohama 226-8502, Japan Japan.}

\begin{abstract}
Since Self-Organised Criticality (SOC) was introduced in 1987, both the nature of the self-organisation and the criticality remains controversial. Recent observations on rain precipitation and brain activity suggest that real systems display a dynamics that is similar to the one observed in SOC systems, making a better understanding of such systems more urgent. Here we focus on the Drossel-Schwable forest-fire model (FFM) of SOC and show that despite the model has been proved to not being critical, it nevertheless exhibits a behavior that justifies the introduction of a new kind of $\textit{weak}$ criticality.
\end{abstract}
\maketitle
\begin{figure}
\includegraphics[width = \linewidth]{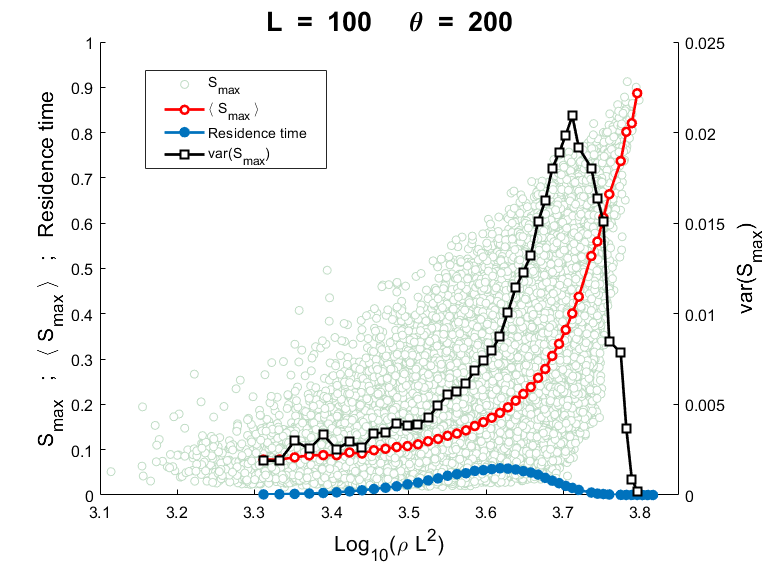}
\caption{Behavior of the control parameter - the size of the biggest cluster normalized to the number of active sites $S_{max}$ - and related observables as a function of the logarithm of the number of active sites. This analysis is equivalent to the one performed in \cite{Chialvo2012} and shows that the behavior of the two systems in the respect is qualitatively the same.}\label{Brain}
\end{figure}
\begin{figure}
\includegraphics[width = \linewidth]{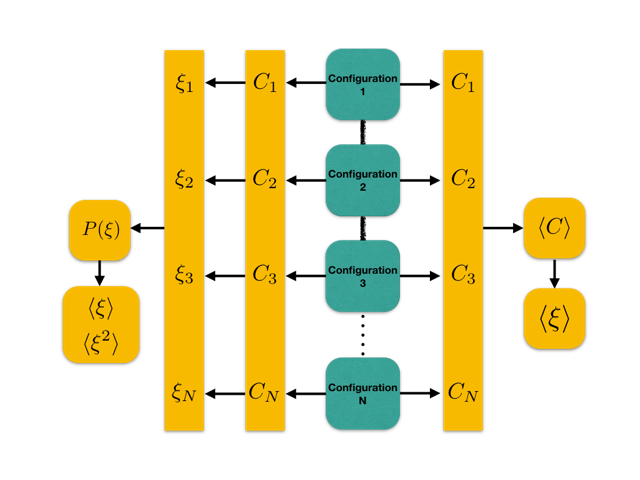}
\caption{Diagram showing two different methods to compute $\langle \xi \rangle$.}\label{diagram}
\end{figure}
\begin{figure}
\includegraphics[width = \linewidth]{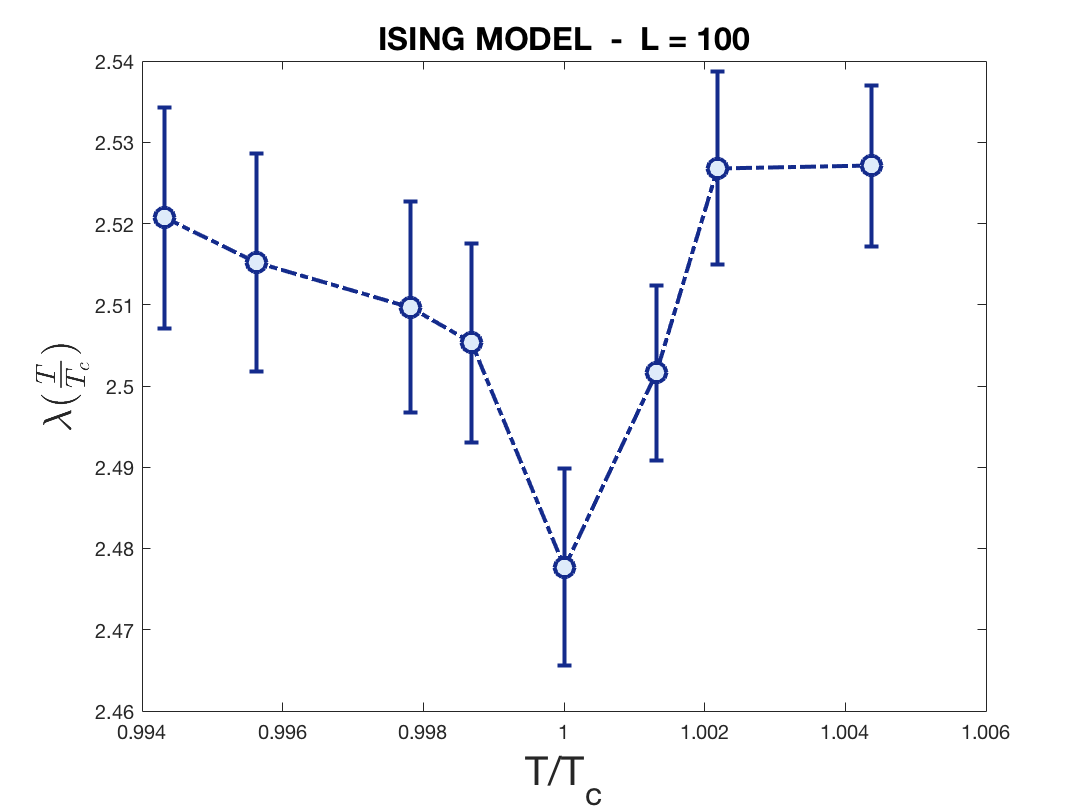}
\caption{This figure represents $\lambda(\frac{T}{T_c})$ for a 2D Ising model with system size $L=100$. For the statistics a sample of $10^6$ data points has been used for each $T$. Error bars are set to $3 \sigma$. }\label{Ising}
\end{figure}

Much of the research inspired by Self-organised Criticality took up the mantle from the paper by Bak, Tang and Wiesenfeld\cite{BTW1987} and studied how various simple dynamical systems may drive themselves into a critical state. Reviews can be found in \cite{Jensen1998,Gunnar_Book,25Years}. We are inspired to return to the discussion concerning the nature of the self tuning to a critical state, or to the vicinity of such a state, by the similarity found when analysing the size distribution of rain showers \cite{Peters_Neelin2006} and the bursts of brain activity measured during fMRI scans\cite{Chialvo2012}. Both studies find indications of critical behavior in terms of approximate power laws and even features reminiscent of peaked, or perhaps diverging, fluctuations or susceptibilities. Also, both studies  investigated the time spent at different values of the control parameter (water vapour and number of voxels activated above threshold respectively), finding that the distribution of residence times, the amount of time spent at a certain value of the control parameter, is found to exhibit a broad peak which indicates that the system spends most of the time near the transition point. The immediate interpretation seems to be that for systems like the atmosphere or the brain, the dynamics consists in some kind of feedback mechanism that is able to bring the system into the vicinity of a critical point. However, the dynamics couples the value of the control parameter to the fluctuations in such a way that even for such big systems fluctuations manage to drive the system away from the critical point. This is similar to suggestions previously put forward such as \cite{Zapperi1995,Sornette1992}. Interestingly, the same behavior can be observed in one of the paradigmatic models of SOC, the Drossel-Schwable Forest Fire Model (FFM) \cite{Drossel1992,Clar1994,Schenk2000,Scott2014}, see for example Fig. \ref{Brain}.  It was very early realised that the FFM doesn't exhibit exact scaling as seen for ordinary equilibrium critical systems\cite{Grassberger1993,Pruessner2002,Grassberger2002} and that this may be related to such feedback dynamics. Hitherto, most people have probably considered the lack of scaling in SOC models (the same complicated scaling is also seen for the original BTW sand-pile model, See Chap. 4 in \cite{Gunnar_Book}, as a sign of broken promise in the sense that SOC suggested that the scenario of criticality and scaling seen in equilibrium systems were generic for a broad range of driven systems. Here we re-analyse the FFM with the observations on precipitation and brain activity in mind and we show that the peculiar dynamics observed both in the FFM and in real systems might be associated to a new kind of \textit{weak} criticality. Although the definition of criticality can be subtle, we will show that our methodology allows to define the degree of criticality of a system based on the way the correlation length diverges.
The classical version of the FFM model consists in a dynamics that involves the occupation of empty sites (birth of new trees) and the removal of randomly selected clusters of sites (when a tree catches fire the whole connected cluster is removed) in a square lattice of size $L^2$. Therefore, there are two different time scales related to the rate of growth of the trees $p$ and the rate of fires $f$. However, in most recent implementations of the model, what is actually used is the ratio between these two rates: $\theta =\frac{p}{f}$. Here we follow the algorithm that is applied also in \cite{Pruessner2002}, and that can be sketched as follows:
\\ \\ FOREVER $\lbrace$ 
\\  REPEAT $\theta$ TIMES $\lbrace$ \\
choose randomly a site $s$; \\
IF( $s$ is empty) THEN $\lbrace$ $s$ becomes occupied $\rbrace $
\\      $\rbrace $
\\ 	choose randomly a site $s$; \\ 
IF( $s$ is occupied) THEN $\lbrace$ 
\\ collect statistics;
\\ burn the whole cluster related to $s$ ;
\\ $\rbrace $
\\ $\rbrace $ 
\\It is important to note that to assure criticality one should use $\theta>>1$. However, the value of $\theta$ is limited by the system size $L$ and should be tuned accordingly to it. 
In order to investigate the properties of the FFM near criticality we will focus on the correlation length $\xi$ as a function of the control parameter of our system, that is the density of occupied sites $\rho$. The classical approach for this kind of analysis consists of computing the two-point correlation function $C(r)$ at different times during the evolution and then obtain the time-averaged correlation function $\langle C(r) \rangle$, for which is assumed the following shape:
\begin{equation}\label{eq: CF}
\langle C(r) \rangle \propto r^{-\eta}e^{-\frac{r}{\xi}}
\end{equation}
Fitting Eq. (\ref{eq: CF}), it is possible to get an estimate of $\xi$. Here we use a different approach. As the system evolves in time through different configurations, we compute a set of correlation functions $\lbrace C_1, C_2 ,C_3 , \ldots ,C_N\rbrace$, each realisation $C_i$ of the correlation function is obtained by a spacial average over a given configuration. We fit each $C_i$ to estimate the correspondent $\xi_i$. From the set of correlation lengths $\lbrace \xi_1, \xi_2 , \xi_3 , \ldots ,\xi_N\rbrace$ we can then obtain the probability density function $P(\xi)$. In this way, we keep the information about the individual configurations instead of averaging them out. Indeed, in the vicinity of a critical point it is possible to encounter specific configurations that either exhibit predominately  power-law or exponentially decaying correlations, i.e. configurations with vastly different values of $\xi_i$. These fluctuations in the range of the correlations are lost when one directly computes the average correlation function and from it extracts the correlation length. Fig. \ref{diagram} illustrates the difference between the usual way of computing $\langle \xi \rangle$ and the procedure we propose.  Once we have $P(\xi)$, we can easily estimate $\langle \xi \rangle$ and further moments of the distribution.
We now explain how the major contribution of this approach lies in the analysis of the tail of the distribution, which we can relate to "degrees of criticality". It turns out that, at least for the FFM and the Ising model, a region of the control parameter $X$ exists for which $P(\xi)$ is a fat-tailed distribution, i.e. $P(\xi) \sim \xi^{-\lambda(X)}$ for large values of $\xi$. It is therefore possible to investigate the presence of diverging or long range correlations just by looking at the behavior of $\lambda(X)$ in the following way. Let us by $X_{critical}$ denote the value of the control parameter $X$ for which correlations extend the farthest. We will call $X_{critical}$  the critical point even if the correlation length, like in the FFM, isn't infinite. Though for the Ising model $X_{critical}=T_c$, the critical temperature.  Introducing 
    $\lambda(X_{critical})=\lambda_c$  at the critical point the integral
\begin{equation}\label{eq: Integral}
I(\alpha )=\int_1^\infty  \xi^{\alpha} P(\xi) d\xi
\end{equation}
is finite for $\alpha - \lambda_c < -1$. This means that the first moment of $P(\xi)$ is finite if $\lambda_c > 2$ and the second if $\lambda_c >3$. Therefore, if $\lambda_c \leq 2$ the mean correlation length diverges and Eq. (\ref{eq: CF}) gives a power-law decaying correlation function at $X_{critical}$.
\\ In order to get some intuition, we can take as a case study the $2D$ Ising model. For an infinitely large system, the 2D Ising model exhibits a phase transition at a fixed value of the control parameter, and $I(1)$ diverges. In other words, we expect that at the critical temperature $T_c$ 
\begin{equation}\label{eq: ising}
\lim_{L \rightarrow \infty} P(\xi \vert  T=T_c) \sim \xi^{-\lambda_c}
\end{equation}
with $\lambda_c=2$. The reason why we expect to have exactly $\lambda_c=2$ in the 2D Ising model is that for any other value $\lambda_c<2$ there would be a range of temperatures close to $T_c$ for which $\langle \xi \rangle$ diverges. Although this could be possible for other systems, it is not the case for the 2D Ising model, where the critical point is unique. For a finite system, we expect $\lambda(T)$ to decrease near $T_c$ and become closer and closer to $2$ as $L$ increases. We have performed this analysis for small sizes $L$ of the Ising model, in order to check wheter there is a minimum for $\lambda(T)$ in correspondence of the critical point and if that minimum is close to $2$. It turns out that even for small systems like $L=50$ it is possible to identify a net drop of $\lambda(T)$ in correspondence of the effective critical temperature. It is important to notice that for finite systems, $T_c$ depends on the system size $L$ \cite{Effective Tc}. Here we use the value that corresponds to the peak in the magnetic susceptibility as the effective $T_c$. Fig. \ref{Ising} shows the results for $L=100$. 

\begin{figure}
\includegraphics[width = \linewidth]{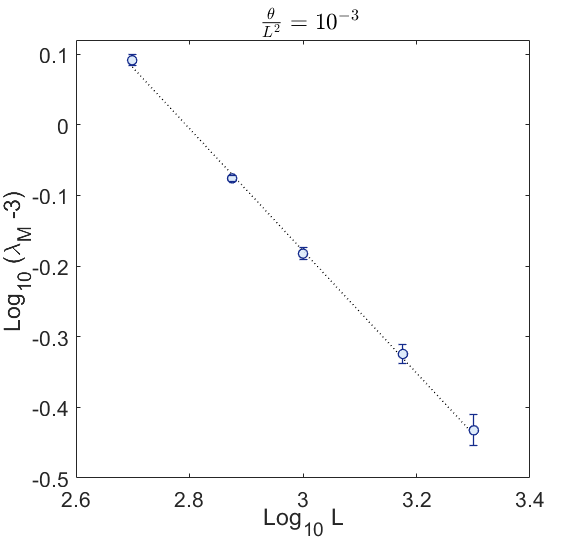}
\caption{Scaling of $\lambda_M - 3 $. For each point a sample of $10^6$ configurations has been used and the ratio $\frac{\theta}{L^2}$ has been kept fixed for each simulation. Error bars are set to $3 \sigma$.}\label{scaling}
\end{figure}

In the case of FFM, the situation is more complicated because the value of $\theta$ has to be tuned with $L$  in order to reach a critical state. Therefore, in the FFM one has a critical surface $\lambda(\theta,\rho)$. From a numerical point of view, is also not trivial to measure accurately the value of the control parameter $\rho$. Since the control parameter is not fixed as in the Ising model, one should define a range $\rho \pm \bigtriangleup \rho$ to collect the statistics needed to compute $\lambda(\rho)$. This range should not be too small in order to have sufficient statistics, but at the same time choosing a large $\bigtriangleup \rho$ would introduce spurious statistics from points close to $\lambda(\rho)$. This problem can be avoided considering the distribution
\begin{equation}\label{eq: Int}
P(\xi) = \int_0^1 P(\xi\mid\rho) P(\rho) d\rho
\end{equation}
Since $P(\rho)$ (see Fig. \ref{Brain}) behaves like a Gaussian distribution with a sharp peak and $P(\xi) \rightarrow \xi^{-\lambda_M}$ for large values of $\xi$, by means of a saddle point approximation we obtain
\begin{equation}\label{LM}
\lambda_M \sim \lambda_c +\frac{1}{2}\frac{d^2 \lambda(\rho)}{d \rho^2}\mid_{\rho_c} (\langle \rho \rangle -\rho_c)^2
\end{equation}
In the large $L$ limit $\langle \rho \rangle \simeq \rho_c$ \cite{Grassberger2002}, therefore $\lambda_M$ tends to $\lambda_c$ as $L$ increases. Furthermore, since the second derivative of $\lambda(\rho)$ is positive near $\rho_c$, we expect $\lambda_M \rightarrow \lambda_c$ from above. Thanks to Eq. (\ref{LM}) it is then possible to have a robust estimate of $\lambda_c$ as the limiting value of $\lambda_M$. Regarding the choice of $\theta$, we have tried different values of $\theta$ for each of the tested values of $L$, and for sufficiently large $\theta$, $\lambda_M$ didn't change noticeably. Fig. \ref{scaling} displays the results for simulations in which the ratio between $L^2$ and $\theta$ has been kept fixed to $\frac{\theta}{L^2}=10^{-3}$. Fitting $\lambda_M(L)$ with a power-law $\lambda_M=\lambda_c + a^{bL}$ we obtained $\lambda_c = 3.10 \pm 0.16 $ and $b=-1.01 \pm 0.3$. In Fig.  \ref{scaling} $\lambda_c$ is assumed to be equal to $3$ and it can be seen that the rescaled variable $\lambda_M-3$ as a function of $L$ is consistent with a straight line in a log-log plot. The analysis of the power laws has been done following the guidelines presented in \cite{PL}. In particular, we have used the Kolmogorov-Smirnov test to identify the beginning of the tail and then the method of maximum likelihood to estimate the scaling exponent. Finally, we have estimated the uncertainties in the estimated power-law exponent by means of the non-parametric approach described in \cite{PL}. Fig. \ref{scaling} is consistent with the ansatz 
\begin{equation}\label{eq: app}
 \lambda_M   \xrightarrow[L \to \infty]{}  3
\end{equation}
Therefore, the FFM displays a completely different scenario compared to the one observed in the Ising model, where $\lambda_c =2$. Indeed, in the Ising model we have that $<\xi>$ diverges at the critical point, while in the FFM the first moment is finite and the second diverges. In other words,  even though the mean correlation length is finite, there is a certain probability of observing configurations with diverging correlation length because of the diverging variance of $P(\xi)$. This could be the reason why the FFM displays certain characteristics of critical systems without however being critical\cite{Pruessner2002} according to the usual definition used in statistical mechanics, namely that $\langle \xi \rangle=\infty$ at the critical point\cite{Christensen_Moloney2005,Ma1985}. In order to distinguish between the classical notion of criticality and the behavior observed in the FFM we will refer to the second as \textit{weak} criticality.
\\ The new analysis method we propose allows us to introduce a quantitative measure of the degree of criticality of a system, i.e. $\lambda_c$, and in particular to distinguish between situations in which one has full criticality like in the Ising model ($\lambda_c=2$) and others in which only certain features of critical systems are preserved, like in the FFM ($\lambda_c=3$). This is particularly relevant for the understanding of the mechanisms that are behind the approximate power laws and other critical features observed in real systems like brain and rain, and to classify the nature of criticality and phase transitions in such systems. Furthermore, this procedure can in principle be easily adapted to experimental or observational settings, since it only requires the analysis of spatial correlations taken from independent configurations of the system at a given value of the control parameter.

LP gratefully acknowledges an EPSRC-Roth scholarship from the Department of Mathematics at Imperial College London, and the High Performance Computing (HPC) facilities provided by the Research Computing Service.

\end{document}